\documentclass[conference]{IEEEtran}
\IEEEoverridecommandlockouts
\usepackage{cite}
\usepackage{amsmath,amssymb,amsfonts}
\usepackage{algorithmic}
\usepackage{graphicx}
\usepackage{textcomp}
\usepackage{xcolor}
\usepackage{enumitem}
\usepackage[misc,geometry]{ifsym}
\usepackage{hyperref}

\usepackage{graphicx}
\usepackage{float}
\usepackage{booktabs} 
\usepackage{threeparttable} 
\usepackage{graphicx} 
\usepackage{booktabs} 
\usepackage{multirow} 
\usepackage{graphicx} 
\def\BibTeX{{\rm B\kern-.05em{\sc i\kern-.025em b}\kern-.08em
    T\kern-.1667em\lower.7ex\hbox{E}\kern-.125emX}}

\begin{document}

\title{ProjectedEx: Enhancing Generation in Explainable AI for Prostate Cancer}

\author{Xuyin Qi$^{1,2,*,\dag}$,
Zeyu Zhang$^{1,3,*,\dag,\ddag}$,
Aaron Berliano Handoko$^{4,*}$,
Huazhan Zheng$^{5}$,
Mingxi Chen$^{6}$,
Ta Duc Huy$^{2}$,\\
Vu Minh Hieu Phan$^{2}$,
Lei Zhang$^{7}$,
Linqi Cheng$^{8}$,
Shiyu Jiang$^{9}$,
Zhiwei Zhang$^{10}$,
Zhibin Liao$^{2}$,\\
Yang Zhao$^{11,\text{\Letter}}$,
Minh-Son To$^{1}$\thanks{$^{*}$Equal contribution. $^{\text{\Letter}}$Corresponding author: \href{mailto:y.zhao2@latrobe.edu.au}{y.zhao2@latrobe.edu.au}}\thanks{$^\dag$Work done while Zeyu Zhang is a student researcher at Flinders University.}\thanks{$^\ddag$Project lead.}
\thanks{$^1$Flinders University $^2$The University of Adelaide $^3$The Australian National University $^4$The University of Sydney $^5$Zhejiang University of Technology $^6$Guangdong Technion – Israel Institute of Technology $^7$University of Chinese Academy of Sciences $^8$Rice University $^9$Johns Hopkins University $^{10}$The Pennsylvania State University $^{11}$La Trobe University}}

\maketitle

\begin{abstract}
Prostate cancer, a growing global health concern, necessitates precise diagnostic tools, with Magnetic Resonance Imaging (MRI) offering high-resolution soft tissue imaging that significantly enhances diagnostic accuracy.
Recent advancements in explainable AI and representation learning have significantly improved prostate cancer diagnosis by enabling automated and precise lesion classification. However, existing explainable AI methods, particularly those based on frameworks like generative adversarial networks (GANs), are predominantly developed for natural image generation, and their application to medical imaging often leads to suboptimal performance due to the unique characteristics and complexity of medical image. To address these challenges, our paper introduces three key contributions. First, we propose \textbf{ProjectedEx}, a generative framework that provides interpretable, multi-attribute explanations, effectively linking medical image features to classifier decisions. Second, we enhance the encoder module by incorporating feature pyramids, which enables multiscale feedback to refine the latent space and improves the quality of generated explanations. Additionally, we conduct comprehensive experiments on both the generator and classifier, demonstrating the clinical relevance and effectiveness of ProjectedEx in enhancing interpretability and supporting the adoption of AI in medical settings. 
Code will be released at \url{https://github.com/Richardqiyi/ProjectedEx}.
\end{abstract}
\section{Introduction}
Prostate cancer is the second most common malignancy in men worldwide, with significant variation in incidence and mortality based on age, race, genetic, social, and environmental factors~\cite{rawla2019epidemiology}.
MRI plays a pivotal role in prostate cancer diagnosis by improving risk stratification and reducing unnecessary biopsies. Multiparametric MRI (mpMRI) combines anatomical and functional imaging, enabling accurate detection of clinically significant cancers while minimizing the overdiagnosis of indolent cases~\cite{tay2021utility}. T2-weighted imaging (T2WI), diffusion-weighted imaging (DWI) with apparent diffusion coefficient (ADC) maps, and T1-weighted dynamic contrast-enhanced (T1DCE) are essential MRI sequences for prostate cancer diagnosis, providing complementary information to detect, localize, and assess the aggressiveness of tumors. T2WI offers detailed anatomical visualization to identify suspicious areas and evaluate capsular invasion, while ADC maps (calculated from DWI) quantify water diffusion, with restricted diffusion appearing as low signal intensity in cancerous regions. DWI enhances tumor visibility by highlighting areas of restricted diffusion with high signal intensity at high b-values, improving contrast between malignant and benign tissues~\cite{tamada2021diffusion,teicua2023tumor,tay2021utility,rosenkrantz2016prostate}. An abbreviated protocol, biparametric MRI (bpMRI), includes T2WI and DWI/ADC only, and is a contrast-free alternative to mpMRI.  

\begin{figure}[t]
    \centering  \includegraphics[width=\columnwidth]{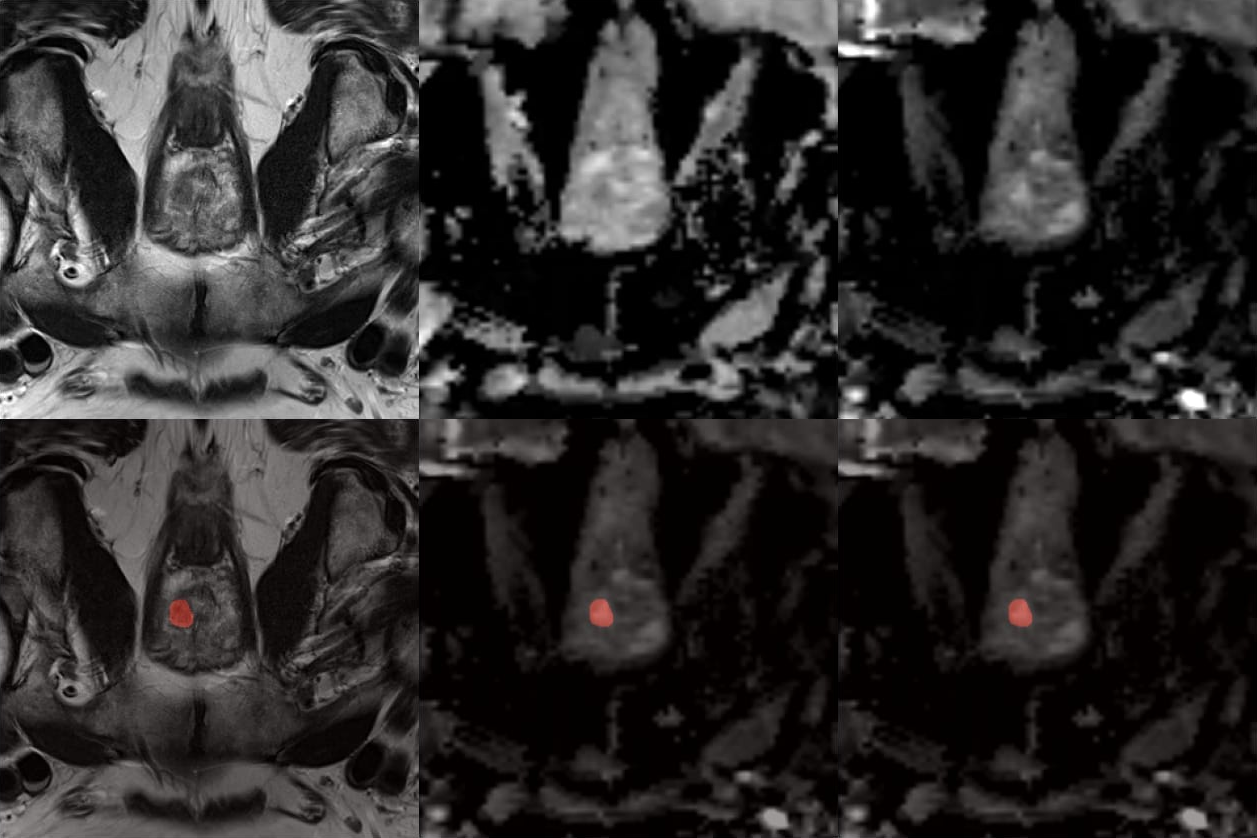} 
    \caption{Visualization of three MRI modalities (T2WI, DWI, ADC) and corresponding prostate cancer locations. The first row shows the original images for each modality, while the second row highlights the prostate cancer regions in red.}
    \label{fig:mri_modalities}
\end{figure}

Data-driven methods have shown significant potential to improve prostate cancer diagnosis by automatically extract complex features from medical images, reducing the reliance on hand-crafted features and domain expertise~\cite{khanfari2023exploring}. However, the lack of interpretability of the models limits clinical adoption, as predictions often lack clear explanations. Developing explainable AI is essential to ensure reliability and trust, particularly in identifying malignancies from MRI data by linking predictions to interpretable features, allowing clinicians to validate and act on diagnostic results~\cite{qian2023recent}.

StylEx~\cite{lang2021explaining} is a framework designed to enhance the interpretability of image classifiers by generating counterfactual explanations based on StyleGAN~\cite{karras2019style}'s StyleSpace. It identifies classifier-specific attributes and visualizes their impact on predictions, offering intuitive, image-specific explanations for decisions. Although effective in natural image classification tasks, such as distinguishing between animal categories, StylEx demonstrates limitations in medical imaging applications. Specifically, its ability to generate realistic and clinically relevant images for MRI data is suboptimal due to the subtlety and complexity of medical features. This highlights the need to develop domain-specific explainable AI models that can produce high-quality, meaningful visualizations tailored to medical imaging, particularly in critical areas like prostate cancer diagnosis.

To address the limitations of current explainability frameworks in medical imaging, we introduce three key contributions:

\begin{itemize}
    \item First, we propose \textbf{ProjectedEx}, a novel method designed to enhance interpretability and trustworthiness. ProjectedEx introduces a generative framework that provides interpretable, multi-attribute explanations by linking specific image features to classifier decisions. This approach bridges the gap between the model outputs and their practical relevance, allowing clinicians to better understand the reasoning behind diagnostic results.
    
    \item Second, we enhance the encoder module by incorporating feature pyramids, which enable multiscale feedback to refine the latent space representations. This architectural improvement significantly enhances the quality, clarity, and relevance of the generated explanations, making them more informative for clinical use.
    
    \item Lastly, we conduct comprehensive experiments to evaluate the performance of the proposed framework. These experiments include assessments of both the generative and classification components, demonstrating that ProjectedEx produces clinically meaningful visualizations and reliable diagnostic insights.
\end{itemize}

\section{Related Work}
Medical imaging analysis \cite{zhao2024landmark,hiwase2024can} has made a significant impact on classification \cite{zhang2024jointvit} and dense prediction \cite{ge2024esa} tasks, including lesion detection \cite{zhang2024meddet,cai2024msdet,cai2024medical} and segmentation \cite{wu2023bhsd,zhang2024segreg,tan2024segstitch,zhang2023thinthick,tan2024segkan}, which assist in diagnosis \cite{zhang2024deep}. Representation learning \cite{ji2024sine,wu2024xlip} has made significant strides in prostate cancer classification, particularly using transfer learning to address limited data challenges. Pre-trained models like InceptionV3~\cite{xia2017inception} and VGG-16~\cite{tammina2019transfer} fine-tuned on the ProstateX dataset have achieved competitive AUCs of 0.81 and 0.83, demonstrating the utility of leveraging ImageNet features~\cite{chen2017transfer}. Automated PI-RADS scoring models, such as fine-tuned ResNet34 pre-trained on ImageNet~\cite{sanford2020deep}, have achieved strong agreement with radiologists. By averaging predictions across slices, these models demonstrated accuracy in lesion-level classification and effectiveness in detecting clinically significant prostate cancers, matching the reliability of human experts. 
However, all these methods lack explanation.

StylEx~\cite{lang2021explaining} is a method that explains image classification decisions by identifying classifier-relevant attributes in StyleGAN's StyleSpace. It modifies these attributes to show their impact on the classifier in generating meaningful, interpretable, and image-specific explanations.

[Re]StylEx \cite{van2021re} then recustomized StylEx in PyTorch to improve reproducibility, addressing missing details from the original code and paper. Experiments confirmed similar results for the pre-trained model but lower accuracy for custom-trained models. The use case validated the coherence of explanations, while FID scores increased with more attributes. Collaboration with the authors helped clarify the training process. 

StylEx is applied to medical imaging to explain chest X-ray classifier decisions. This paper~\cite{atad2022chexplaining} customizes StyleGAN to identify important features and generate counterfactual images by modifying them. The EigenFind algorithm reduces computation time, and the results show clinically relevant insights, helping improve models. Generative models like StyleGAN2\cite{karras2020analyzing} also have demonstrated the ability to capture prostate cancer features in latent spaces without supervision. By training on a prostate histology dataset, the model accurately mapped diagnostic features, with 77\% of synthetic images retaining the same annotations and 98\% matching the same or adjacent diagnostic stages, confirming its potential for modeling prostate cancer pathology~\cite{daroach2022prostate}.

\section{Methods}

\subsection{Overview}
Our proposed framework, ProjectedEx, integrates generative modeling and multiscale feature analysis to link medical image features with classifier decisions. The framework comprises an encoder-decoder structure enhanced by a feature pyramid module and a set of discriminators operating at multiple resolutions. This design ensures the generation of interpretable, multiscale explanations that highlight the connection between image features and classifier outputs.

\begin{figure}[htbp]
    \centering
    \includegraphics[width=\columnwidth]{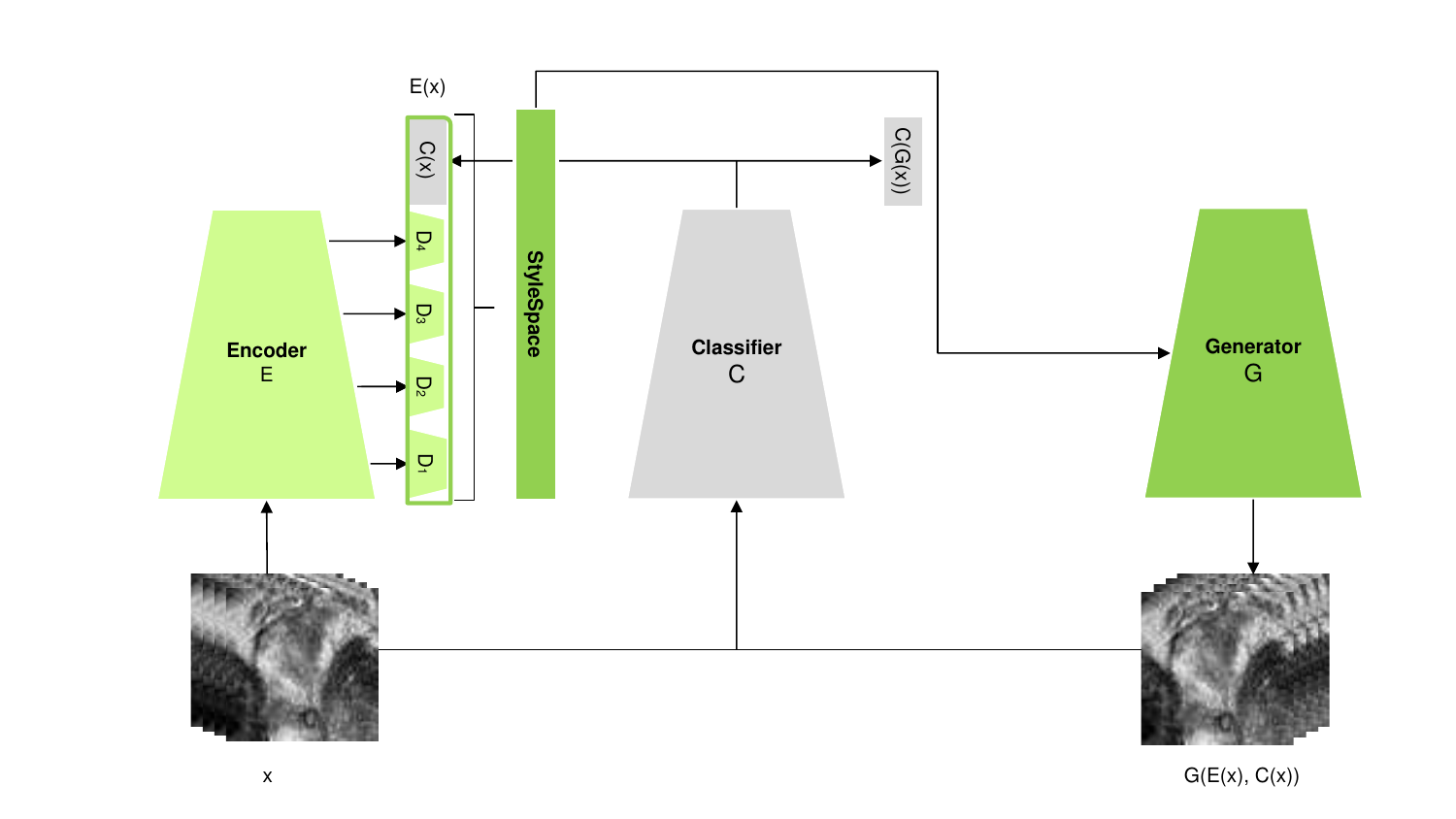}
    \caption{The architecture diagram illustrating the interactions between the encoder (\(E\)), generator (\(G\)), and classifier (\(C\)). Specifically, \(D_1, D_2, D_3, D_4\) are concatenated to form \(E(x)\), and \(x'\) is generated as \(G(E(x), C(x))\). The reconstruction loss \(L_{\text{rec}}^x\) is calculated between \(x\) and \(x'\), while the classification loss \(L_{\text{cls}}\) is computed between \(C(x)\) and \(C(G(x))\). }
    \label{fig:architecture}
\end{figure}

\subsection{Feature Pyramid Encoder}
To capture features across multiple scales, we introduce a feature pyramid encoder that extracts representations from four layers of the encoder, denoted as $L_l$ with resolutions progressively down-sampled to $L_1 = 64^2$, $L_2 = 32^2$, $L_3 = 16^2$, and $L_4 = 8^2$. Each layer’s features are processed by a corresponding discriminator $D_l$, which operates at the same scale.

Each discriminator $D_l$ is designed with a lightweight convolutional architecture and employs spectral normalization to ensure stable training. The outputs of all discriminators are unified at a fixed resolution of $128$, achieved by reducing the number of down-sampling blocks for lower-resolution inputs. The discriminator losses are aggregated to provide feedback for the generator, enabling consistent outputs across scales.

\begin{figure}[ht]
    \centering
    \includegraphics[width=\columnwidth]{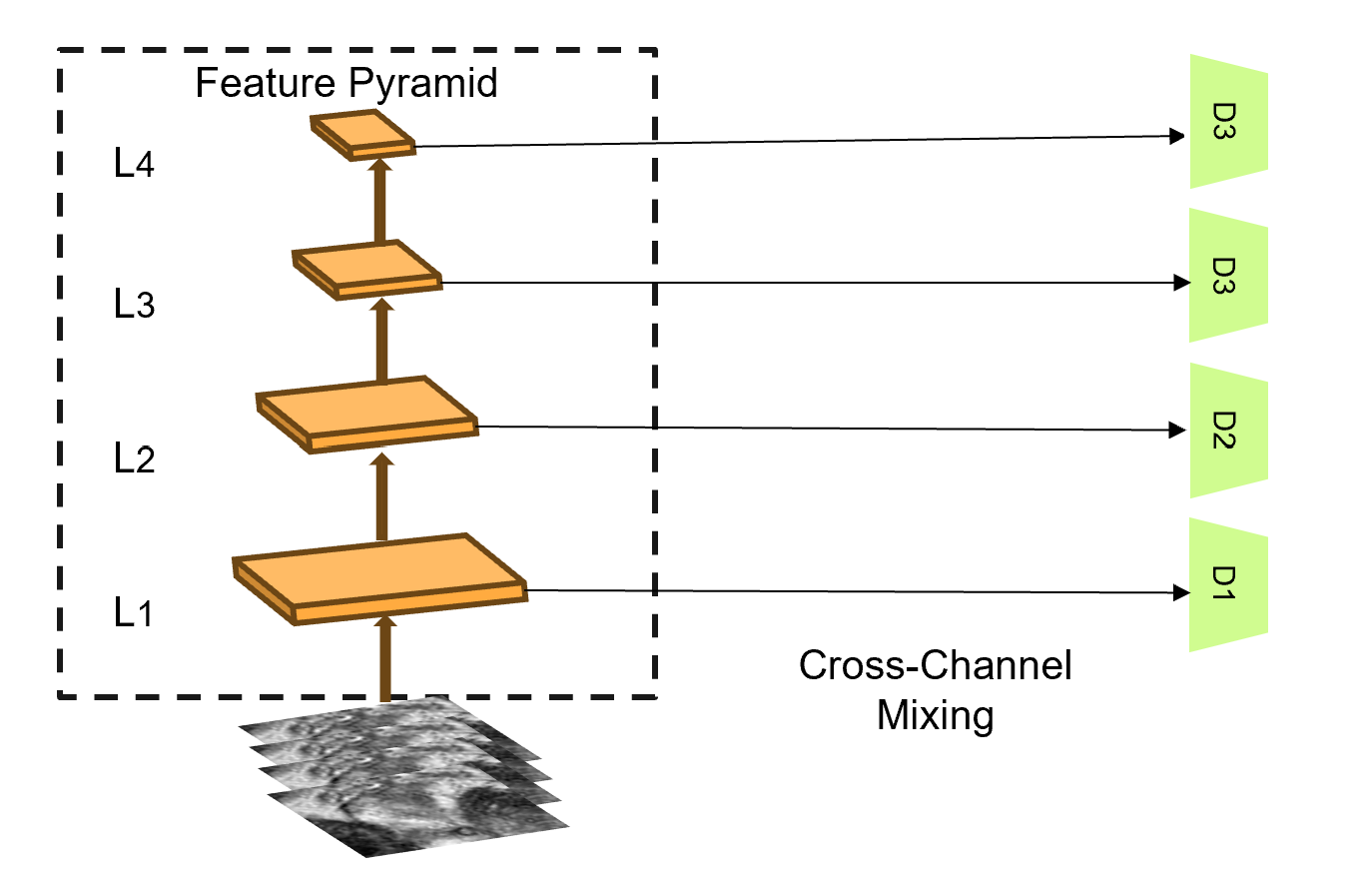} 
    \caption{
        A feature pyramid encoder extracts multiscale features from four layers ($L_1$ to $L_4$), processed by lightweight discriminators ($D_1$ to $D_4$) with spectral normalization. Outputs are unified at a fixed resolution of $128$ for consistent feedback across scales.
    }
    \label{fig:feature_pyramid}
\end{figure}

\subsection{Differentiable Random Projections}
To enhance interpretability and robustness, we incorporate differentiable random projections via Cross-Channel Mixing (CCM). This involves applying a randomly initialized $1 \times 1$ convolution to features at each scale, effectively mixing feature channels while preserving the overall information. Unlike previous methods that rely on rotation matrices for projection, our approach initializes the convolutional weights using Kaiming initialization~\cite{he2015delving}, which promotes effective feature mixing without trivial invertibility.

The projected features are fed into the corresponding discriminators, ensuring that the model effectively leverages information from all scales. This design encourages balanced utilization of multiscale features, leading to a refined latent space and improved generation of interpretable explanations.

\subsection{Goals of Optimization}

\textbf{StyleSpace Construction}
The outputs of all discriminators are concatenated to form a feature vector of 512 dimensions. This feature vector is then combined with the two logits from the classifier, resulting in a 514-dimensional representation referred to as the StyleSpace. By providing the generator with the intended classifier output values for the generated images, this conditioning mechanism ensures that the StyleSpace captures attributes relevant to the classifier’s decision-making process. As a result, the coordinates in the StyleSpace become an affine transformation of the classifier output, embedding more attributes that influence the classifier's predictions.

\textbf{Loss Function}
The overall training loss $L$ combines several components:
\begin{align*}
L = L_{\text{adv}} + L_{\text{reg}} + L_{\text{rec}} + L_{\text{cls}},
\end{align*}
where $L_{\text{adv}}$ is the adversarial loss~\cite{goodfellow2014generative}, and $L_{\text{reg}}$ is the path regularization~\cite{karras2020analyzing}. The reconstruction loss, $L_{\text{rec}}$, includes three terms:
$L^{x}_{\text{rec}} = ||x' - x||_1$, $L_{\text{LPIPS}}$~\cite{zhang2018unreasonable}  \text{and}  $L^{w}_{\text{rec}} = ||E(x') - E(x)||_1$, 
where $x'$ is the reconstructed image. The classifier loss is given by:
\begin{align*}
L_{\text{cls}} = D_{\text{KL}}[C(x') | C(x)],
\end{align*}
ensuring the generated images retain attributes critical for classification.

\subsection{Extracting Attributes}
After training the generative model, the next step is to identify coordinates in the StyleSpace that correspond to classifier-specific attributes. These coordinates represent directions in the latent space where adjustments result in changes to the generated image that significantly affect the classifier output. By modifying these coordinates, counterfactual explanations can be generated to explore how specific features influence the classifier's decision.

To achieve this, a set of images is selected where the classifier predicts labels other than the target class. For each target class, the method identifies a subset of StyleSpace coordinates that, when adjusted, increase the probability of the target class. Each coordinate is assigned a direction indicating whether increasing or decreasing its value impacts the classifier output. The process iteratively evaluates the effect of each coordinate on the classifier's predictions, retaining only those that produce meaningful changes. This approach ensures the discovered attributes are directly linked to the classifier’s decision-making process.

\section{Experiments}
\subsection{Dataset and Evaluation Matrices}

We utilized the publicly available Prostate Imaging: Cancer AI (PI-CAI) Challenge dataset, which provides a comprehensive benchmark for evaluating prostate cancer classification models. The performance of classifiers was evaluated using metrics such as Accuracy, Precision, Recall, and F1 score, while the performance of generative networks was assessed using the Fréchet Inception Distance (FID).

To define risk categories for prostate cancer, we adopted the International Society of Urological Pathology (ISUP) Grade Group system~\cite{egevad2016international}, which is a widely accepted and simplified grading system based on prostate biopsy samples. The ISUP Grade is considered an accurate predictor of how aggressively the cancer might spread. It categorizes tumors into five grade groups, where a higher grade indicates a higher likelihood of aggressive and rapidly spreading cancer. The table below provides the mapping between the Gleason score and the ISUP Grade Group

\begin{table}[h]
    \centering
    \caption{Mapping between ISUP Grade Group, Gleason Score, and associated risk levels. Higher ISUP Grade Groups correspond to increased risk of aggressive and rapidly spreading cancer.}
    \label{tab:isup_grade}
    \resizebox{\columnwidth}{!}{%
        \begin{tabular}{ccp{4cm}}
            \hline
            \textbf{Grade Group} & \textbf{Gleason Score} & \textbf{Risk} \\
            \hline
            1 & 3 + 3 = 6 & Low risk: the cancer is usually slow growing and less likely to spread. \\
            \hline
            2 & 3 + 4 = 7 & Intermediate favourable risk: the cancer can be moderately likely to spread. \\
            \hline
            3 & 4 + 3 = 7 & Intermediate unfavourable risk: the cancer can be moderately likely to spread. \\
            \hline
            4 & 4 + 4 = 8 & High risk: the cancer can be fast growing and more likely to spread. \\
            \hline
            5 & 9 or 10 & The highest risk: the cancer can be fast growing and most likely to spread. \\
            \hline
        \end{tabular}%
    }
\end{table}

In this study, ISUP Grade Groups 1 and 2 (corresponding to Gleason scores $\leq 3+4=7$) were categorized as low-risk, while ISUP Grade Groups greater than 2 (corresponding to Gleason scores $\geq 4+3=7$) were considered high-risk. This binary classification scheme was used to train and evaluate our models, ensuring a clear distinction between low-risk and high-risk prostate cancer cases.

\subsection{Implementation Details}

The proposed ProjectedEx model was implemented using the PyTorch 2.0 framework and trained on an NVIDIA GPU. The training process employed a batch size of 16 and spanned 100,000 steps. To optimize the learning rate during training, we used a cosine decay learning rate scheduler. This approach starts with an initial learning rate, gradually decreases it following a cosine curve, and allows the model to converge more effectively by avoiding rapid learning rate changes.

 The training of the ProjectedEx model began with a pre-training phase, where a classifier was also implemented using the PyTorch 2.0 framework. This classifier was trained with a batch size of 64 for a total of 50 epochs. The dataset was split into an 8:2 ratio for training and testing, ensuring a robust evaluation of the classifier's performance.

 \subsection{Comparative Studies}

The data processing for this experiment involved setting the three MRI modalities (DWI, T2WI, ADC) as the threes channels as an input image. Images were cropped to a size of $64\times64$ centered on the manually annotated mask centroid.

The classification performance of three classifiers: MobileNetv2~\cite{sandler2018mobilenetv2}, ResNet18~\cite{he2016deep}, ShuffleNetv2~\cite{ma2018shufflenet} and EfficientNet-B\cite{tan2019efficientnet}, evaluated based on accuracy, precision, recall, and F1 score. Among the classifiers, EfficientNet-B demonstrated the highest accuracy at 83.97\%, significantly outperforming MobileNet (57.90\%), ResNet18 (61.62\%) and ShuffleNetv2 (64.28\%). This result indicates that EfficientNet-B is the most effective classifier for this task, achieving superior performance in terms of overall accuracy.

\begin{table}[t]
  \begin{center}
    \caption{Performance of Classifiers} \label{tab:cap}
    \resizebox{\columnwidth}{!}{%
    \begin{tabular}{ccccc}
      \hline
      Classifier & Accuracy (\%) & Precision (\%) & Recall (\%) & F1 (\%)
      \\
      \hline
      MobileNetv2     & 57.90     & 58.26 &55.32 &56.75       \\
      ResNet18     & 61.62     & 68.05 &  43.62 &53.16      \\
      ShuffleNetv2   & 64.28 & 69.31 & 51.06 & 58.81       \\
      EfficientNet-B & 83.97 & 91.90 & 90.63 & 91.26       
      \\
      \hline
    \end{tabular}%
    }
  \end{center}
\end{table}

Table II presents a comparative evaluation of different classifiers (MobileNetv2, ResNet18, ShuffleNetv2 and EfficientNet-B) paired with two models, [Re]StylEx and ProjectedEx, using the Fréchet Inception Distance (FID) as the performance metric. The FID is a widely recognized standard for measuring the quality of generated images, where lower scores indicate better performance.

\begin{table}[t]
  \begin{center}
    \caption{Evaluation of FID Performance Across Classifiers for [Re]StylEx and ProjectedEx Models} \label{tab:cap}
    \begin{tabular}{ccc}
      \hline
      Classifier & Model & FID \\
      \hline
      MobileNetv2  & [Re]StylEx & 178.21 \\ 
                   & ProjectedEx & 117.78 \\
      \hline
      ResNet18     & [Re]StylEx & 134.20 \\ 
                   & ProjectedEx & 113.42 \\
      \hline
      ShuffleNetv2 & [Re]StylEx & 149.29 \\ 
                   & ProjectedEx & 144.71 \\
      \hline
      EfficientNet-B & [Re]StylEx & 127.08 \\ 
                     & ProjectedEx & 108.63 \\
      \hline
    \end{tabular}%
  \end{center}
\end{table}

ProjectedEx consistently outperforms [Re]StylEx across all classifiers by achieving significantly lower FID scores, as demonstrated in Table II. For instance, when paired with MobileNetv2, ProjectedEx achieves an FID of 117.78, marking a substantial improvement over [Re]StylEx's 178.21. Similarly, for ResNet18, ProjectedEx outperforms [Re]StylEx by achieving an FID of 113.42 compared to 134.20. ShuffleNetv2 shows a consistent trend, where ProjectedEx attains an FID of 144.71, outperforming the 149.29 of [Re]StylEx. The most notable improvement is observed with EfficientNet-B, where ProjectedEx achieves the lowest FID of 108.63, significantly better than the 127.08 achieved by [Re]StylEx.

\begin{figure}[htbp]
    \centering
    \includegraphics[width=\columnwidth]{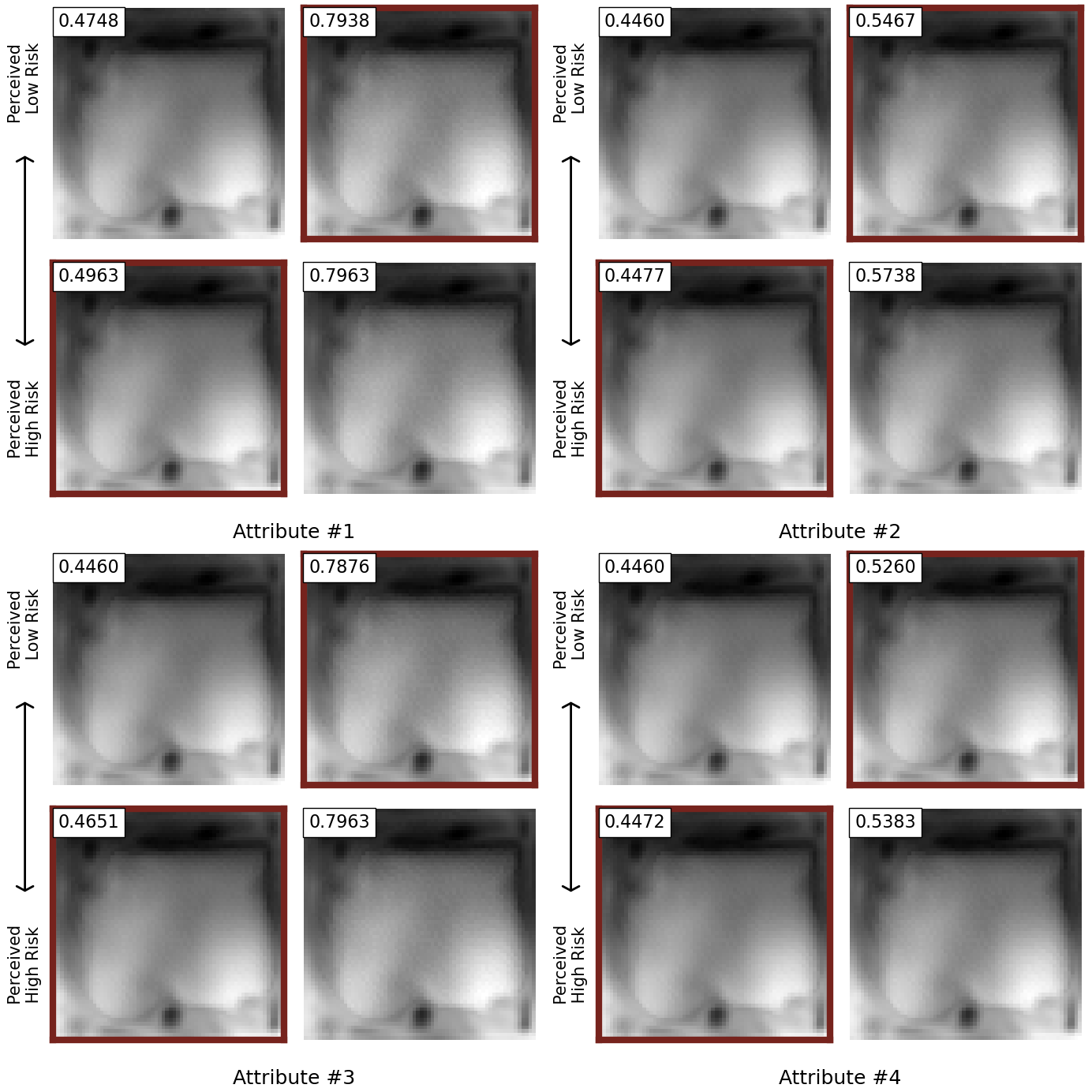}
    \caption{Visualization of the effect of attributes on classifier logits (DWI).}
    \label{fig:dwi_attributes}
\end{figure}

These results strongly validate the performance enhancements brought by ProjectedEx. Across all classifiers, ProjectedEx consistently demonstrates robustness and a superior ability to improve image generation quality. The FID of 108.63 achieved with EfficientNet-B underscores the exceptional capabilities of ProjectedEx, establishing it as the optimal configuration in this experimental setup. The consistent performance improvement across classifiers provides compelling evidence of ProjectedEx's effectiveness and adaptability.

\subsection{Visualization}

In this section, we demonstrate how adjusting attributes can influence the classifier's output results. As illustrated in the example, the left and right columns represent two categories: low-risk prostate cancer and high-risk prostate cancer, respectively. The numbers in the top-right corner of each example indicate the logit outputs of the classifier. Images outlined in red represent those generated after modifying the corresponding attributes. Fig. 4, Fig. 5, and Fig. 6 illustrate the scenarios for DWI, T2WI, and ADC modalities, respectively. 
Attribute 1 to 4 refer to the top four coordinates that have the greatest impact on the classifier's output results.

The results show that by adjusting specific attributes-represented as coordinates in the StyleSpace, it is possible to increase the logits of low-risk category images, making them more likely to be classified as high-risk by the classifier. Conversely, similar adjustments can decrease the logits of high-risk category images, making them more likely to be classified as low-risk. This visualization highlights the impact of targeted attribute manipulation on the classifier's decision-making process.

\begin{figure}[t]
    \centering
    \includegraphics[width=\columnwidth]{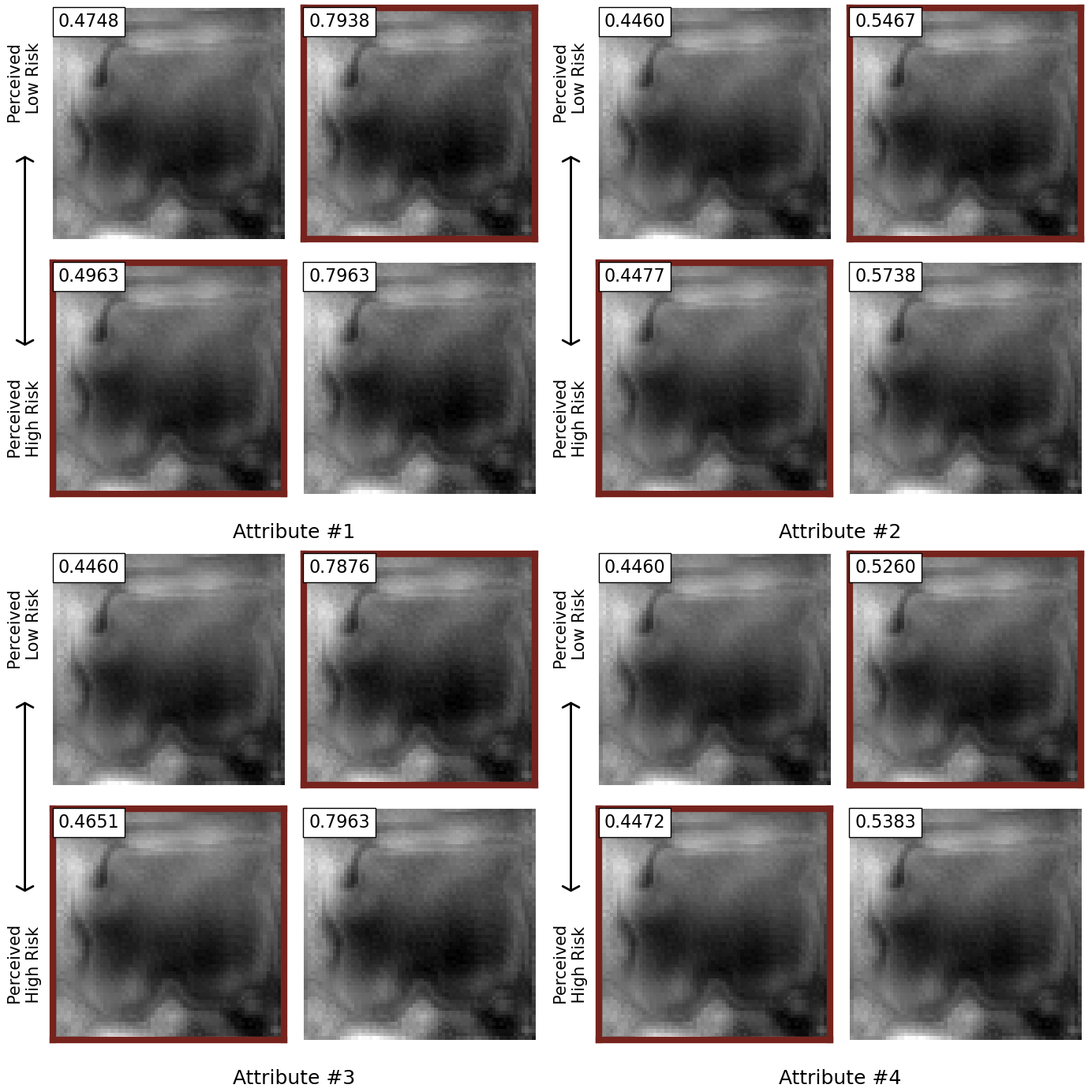}
    \caption{Visualization of the effect of attributes on classifier logits (T2WI).}
    \label{fig:dwi_attributes}
\end{figure}

\subsection{Discussion}

The proposed approach has demonstrated significant advancements in both clinical impact and model performance, offering valuable insights into prostate cancer classification and generative modeling.

\subsubsection{Clinical Impact}
The ability to adjust attributes in StyleSpace and influence classifier outputs has important clinical implications. By identifying how specific attributes correspond to changes in classifier logits, this method provides a novel approach for understanding the underlying features associated with low-risk and high-risk prostate cancer. This capability could aid clinicians in better interpreting AI-driven predictions, ensuring more transparent and explainable decision-making processes. Additionally, this framework allows for the exploration of hypothetical scenarios, such as simulating feature adjustments to assess how certain changes might impact diagnostic outcomes. Such insights could prove invaluable in refining treatment strategies and improving patient care.

\begin{figure}[t]
    \centering
    \includegraphics[width=\columnwidth]{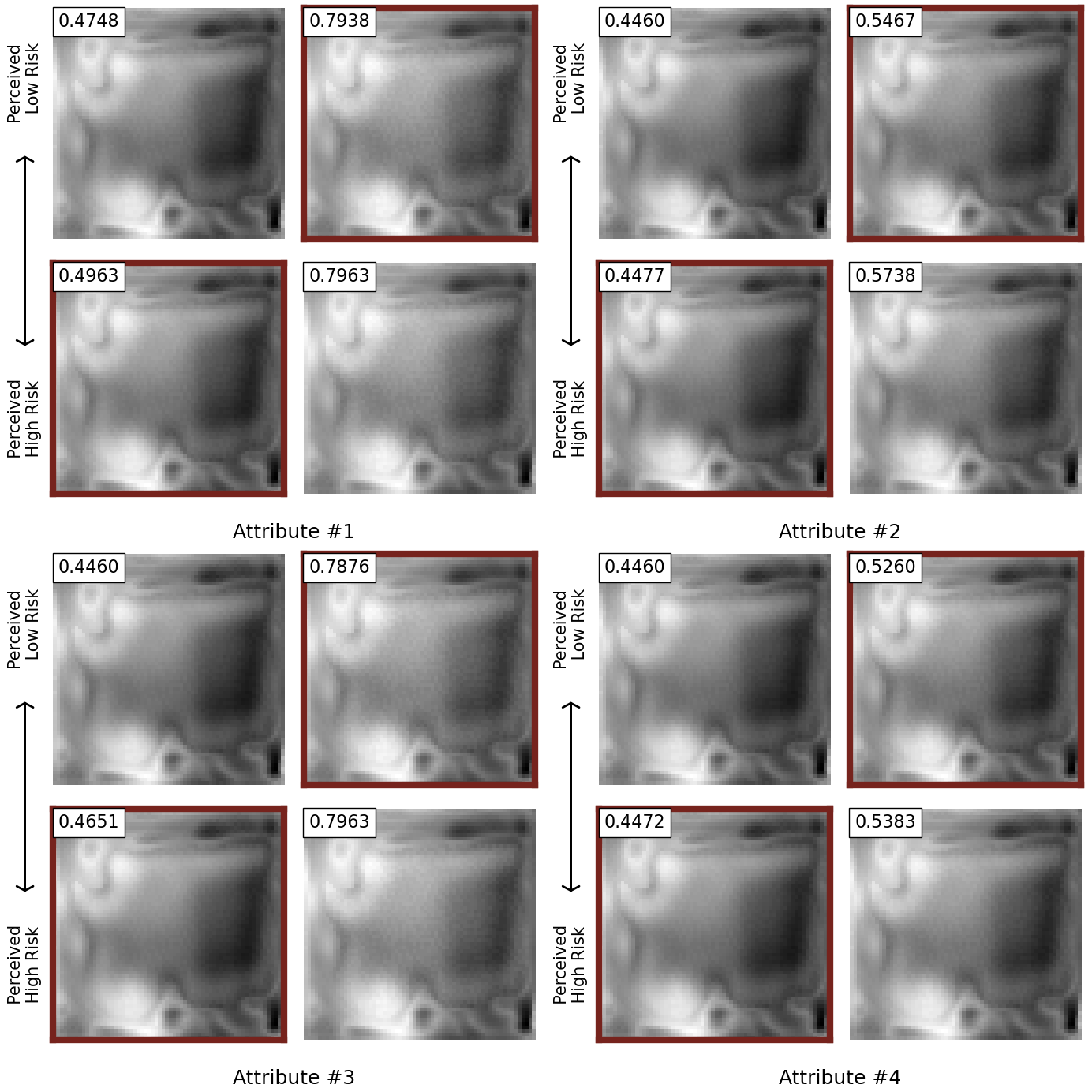}
    \caption{Visualization of the effect of attributes on classifier logits (ADC).}
    \label{fig:dwi_attributes}
\end{figure}

\subsubsection{Model Performance Improvements}
From a performance perspective, the ProjectedEx model consistently outperforms [Re]StylEx across all classifiers in terms of FID, showcasing its superior generative capabilities. The ability to generate high-quality images with improved realism ensures more robust downstream applications, including classifier training and visualization. Furthermore, the EfficientNet-B classifier demonstrated the best accuracy, precision, recall, and F1 score, highlighting the importance of selecting optimal architectures for prostate cancer classification tasks. The significant reduction in FID when using ProjectedEx, particularly with EfficientNet-B (108.63), underscores the effectiveness of combining robust classifiers with advanced generative models to achieve state-of-the-art performance.

Overall, the combination of enhanced generative modeling and classification accuracy not only advances the technical landscape but also lays the groundwork for practical applications in clinical settings, bridging the gap between model-driven insights and actionable medical decisions.

\section{Conclusion}

This paper presents a novel framework, ProjectedEx, for prostate cancer classification and explanation in medical imaging, specifically designed to address challenges in interpretability and multiscale feature representation. To enhance the alignment between classifier outputs and generative explanations, we introduced a Feature Pyramid Encoder, which captures multi-resolution features critical for understanding complex image patterns. Additionally, we employed Differentiable Random Projections via Cross-Channel Mixing (CCM) to improve the robustness of feature representations across scales while maintaining model interpretability.

To link medical image features with classifier decisions, we proposed a StyleSpace construction mechanism that encodes attributes influencing classifier outputs as affine transformations in the latent space. This enables precise manipulation of StyleSpace coordinates to explore how specific attributes impact classification logits, providing direct insight into the features driving predictions. Furthermore, we integrated an advanced loss function combining adversarial, reconstruction, and classification losses to ensure the generation of high-quality, attribute-sensitive counterfactual images.

Empirical validation on the public PI-CAI dataset demonstrates that ProjectedEx achieves state-of-the-art results in both generative quality and classification accuracy. The model significantly reduces Fréchet Inception Distance (FID), with the best score of 108.63 when paired with EfficientNet-B, and achieves the highest classification accuracy of 83.97\%. These results underscore the effectiveness of ProjectedEx in both enhancing the interpretability of data-driven predictions and improving the reliability of prostate cancer diagnosis.

\end{document}